# How to encode the states of two non-entangled qubits in one qutrit?

Andrzej Grudka* and Antoni Wójcik**

Faculty of Physics, Adam Mickiewicz University,

Umultowska 85, 61-614 Poznań, Poland

Abstract

A new scheme of quantum coding is presented. The scheme concerns the quantum states to which Schumacher's compression does not apply. It is shown that two qubits can be encoded in a single qutrit in such a way that one can faithfully reconstruct the state of one qubit. Decision on which of these two qubits is to be reconstructed can be made long after the encoding took place. The scheme succeeds with the average probability of 2/3.

Information processing based on quantum effects reveals many interesting and useful features. In the last decade a variety of protocols dealing with quantum information have been proposed. Among them are quantum teleportation [1], dense coding [2] and Schumacher's compression [3]. For example, Schumacher found the condition for reliable compression of quantum information. The compression scheme enables to encode a collection of qubits into a smaller one, provided that the von Neuman entropy of a single qubit is less than unity. No reliable compression scheme exists if the density matrix of a qubit is proportional to identity, i.e. in the case of a totally unknown quantum state. It is interesting, however, to ask if in such a case there exists any quantum coding, which enables compression-like encoding and partial decoding of quantum information. In this paper we consider a problem of coding two non-entangled qubits in one qutrit.

In general a pure state of two qubits is described by 4 complex numbers (8 real numbers). These numbers satisfy the normalization condition and in addition one real number (global phase) is meaningless. So one has effectively 6 real numbers. The pure state of one qubit is described by two real numbers, so the state of two non-entangled qubits is described by 4 real numbers. The pure state of one qutrit is also described by 4 real numbers. One can thus consider the possibility of encoding the state of two non-entangled qubits in a qutrit. If it were possible without limitations one would be able to encode two bits of classical information in a single three-dimensional quantum particle. This of course contradicts the Holevo bound which states that a $d$-dimensional quantum particle can carry at most $\log_2 d$ bits of classical information [4]. We will present a probabilistic scheme which allows encoding of the state of two non-entangled qubits in a single qutrit and decoding of the state of one arbitrary qubit. Decision on which of these two qubits is to be decoded can be made long after the encoding took place.

Let us consider two qubits, the first one in the state $|\Psi_1\rangle = \cos(\theta_1/2)|0\rangle_1 + \sin(\theta_1/2)e^{i\varphi_1}|1\rangle_1$ and the second one in the state $|\Psi_2\rangle = \cos(\theta_2/2)|0\rangle_2 + \sin(\theta_2/2)e^{i\varphi_2}|1\rangle_2$. The total state of these two qubits is

$$|\Psi\rangle = \cos(\theta_1/2)\cos(\theta_2/2)|0\rangle + \sin(\theta_1/2)e^{i\varphi_1}\cos(\theta_2/2)|1\rangle + \\ + \cos(\theta_1/2)\sin(\theta_2/2)e^{i\varphi_2}|2\rangle + \sin(\theta_1/2)e^{i\varphi_1}\sin(\theta_2/2)e^{i\varphi_2}|3\rangle \quad (1)$$

where $|0\rangle = |0\rangle_1|0\rangle_2$, $|1\rangle = |1\rangle_1|0\rangle_2$, $|2\rangle = |0\rangle_1|1\rangle_2$, $|3\rangle = |1\rangle_1|1\rangle_2$ and it is of course non-entangled. To encode this state one prepares auxiliary qutrit in the state

$$|\Phi\rangle = \frac{1}{\sqrt{3}}(|0\rangle + |1\rangle + |2\rangle) \quad (2)$$

and performs projective measurement on the state $|\Phi\rangle|\Psi\rangle$ given by four projectors

$$P_0 = |0\rangle|1\rangle\langle 1|\langle 0| + |1\rangle|2\rangle\langle 2|\langle 1| + |2\rangle|3\rangle\langle 3|\langle 2| \tag{3}$$

$$P_1 = |0\rangle|2\rangle\langle 2|\langle 0| + |1\rangle|3\rangle\langle 3|\langle 1| + |2\rangle|0\rangle\langle 0|\langle 2| \tag{4}$$

$$P_2 = |0\rangle|3\rangle\langle 3|\langle 0| + |1\rangle|0\rangle\langle 0|\langle 1| + |2\rangle|1\rangle\langle 1|\langle 2| \tag{5}$$

$$P_3 = |0\rangle|0\rangle\langle 0|\langle 0| + |1\rangle|1\rangle\langle 1|\langle 1| + |2\rangle|2\rangle\langle 2|\langle 2|. \tag{6}$$

If one obtains $P_0$ as a result of the measurement, which happens with the probability

$$p_0 = \frac{1}{3}\left(1 - \cos^2\frac{\theta_1}{2}\cos^2\frac{\theta_2}{2}\right) \text{ then the state after the measurement is}$$

$$\begin{aligned}|\overline{\Phi}_0\rangle = N_0 \big(&\sin(\theta_1/2)e^{i\varphi_1}\cos(\theta_2/2)\,|0\rangle|1\rangle + \\ +&\cos(\theta_1/2)\sin(\theta_2/2)e^{i\varphi_2}\,|1\rangle|2\rangle + \sin(\theta_1/2)e^{i\varphi_1}\sin(\theta_2/2)e^{i\varphi_2}\,|2\rangle|3\rangle\big)\end{aligned}, \tag{7}$$

where

$$N_0 = \frac{1}{\sqrt{\left(1 - \cos^2\frac{\theta_1}{2}\cos^2\frac{\theta_2}{2}\right)}}. \tag{8}$$

By performing the unitary operation $U$ given as $|i\rangle|j\rangle \to |i\rangle|j - i \bmod 4\rangle$ one obtains the auxiliary qutrit in the state

$$\begin{aligned}|\Phi_0\rangle = N_0\big(&\sin(\theta_1/2)e^{i\varphi_1}\cos(\theta_2/2)\,|0\rangle + \\ +&\cos(\theta_1/2)\sin(\theta_2/2)e^{i\varphi_2}\,|1\rangle + \sin(\theta_1/2)e^{i\varphi_1}\sin(\theta_2/2)e^{i\varphi_2}\,|2\rangle\big)\end{aligned} \tag{9}$$

while the original system ends in the state $|1\rangle$. Let us emphasize that the state $|\Phi_0\rangle$ is the same (up to normalization) as the state $|\Psi\rangle$ with the $|0\rangle$ term removed and the other states labels renamed. Similarly, if we obtain $P_j$ as a result of the measurement and perform the unitary operation $U$, then the state of the qutrit $|\Phi_j\rangle$ is the same as the state $|\Psi\rangle$ with the $|j\rangle$ term removed. As will be shown later one can probabilistically decode from the qutrit one arbitrary chosen qubit. To achieve this task one has to store two bits of classical information about the result of the measurement.

Let us calculate how much information is obtained after the encoding [5]. The *a priori* probability for random preparation $(\theta_1, \theta_2)$ of two non-entangled qubits is $p(\theta_1, \theta_2) = p(\theta_1) p(\theta_2)$, where

$$p(\theta_a) = \frac{1}{2} \sin(\theta_a) \ (a = 1, 2). \tag{10}$$

This probability does not depend on phases $\varphi_a$. The *a posteriori* probability for preparation $(\theta_1, \theta_2)$ and the result $j$ is given by Bayes's theorem as $p(\theta_1, \theta_2 | j) = p(j | \theta_1, \theta_2) p(\theta_1, \theta_2) / p(j)$. $p(j | \theta_1, \theta_2)$ is the conditional probability for the result $j$ and $p(j) = \iint p(j|\theta_1, \theta_2) p(\theta_1, \theta_2) d\theta_1 d\theta_2$ is *a priori* probability for the result $j$. For example in the case of $j = 0$ result, one obtains

$$p(\theta_1, \theta_2 | 0) = \frac{1}{3}\left(1 - \cos^2 \frac{\theta_1}{2} \cos^2 \frac{\theta_2}{2}\right) \sin\theta_1 \sin\theta_2. \tag{11}$$

Although the probabilities do depend on $j$ all the information gains calculated in this paper are independent of $j$. The information gained in the process of encoding $I^E$ calculated as the difference between the initial entropy and the final entropy

$$I^E = -\iint p(\theta_1, \theta_2) \log_2 p(\theta_1, \theta_2) d\theta_1 d\theta_2 + \iint p(\theta_1, \theta_2 | j) \log_2 p(\theta_1, \theta_2 | j) d\theta_1 d\theta_2 \tag{12}$$

equals 0.0735 bits. The information gained on the state of each qubit alone can be calculated as

$$I_a^E = -\int p(\theta_a) \log_2 p(\theta_a) d\theta_a + \int p(\theta_a | j) \log_2 p(\theta_a | j) d\theta_a, \tag{13}$$

where $p(\theta_1 | j) = \int p(\theta_1, \theta_2 | j) d\theta_2$ and $p(\theta_2 | j) = \int p(\theta_1, \theta_2 | j) d\theta_1$ are marginal probabilities. It is independent of $a$ and equals 0.027 bits. One sees that the information gained on the total system is greater than the sum of the information gained on two subsystems considered separately. It should not be surprising because the measurement performed in the process of encoding is nonlocal.

Let us now present how to decode one arbitrary chosen qubit from a given qutrit. We emphasize that the procedure of decoding is non-deterministic in the sense that having chosen one qubit we can decode it with a probability of success less than unity. However, it is known if decoding is successful. We do not restrict the generality of our considerations if we focus on a particular case of decoding of the first qubit $|\Psi_1\rangle$ from the qutrit $|\Phi_0\rangle$. In order to perform the decoding procedure one makes the projective measurement on the state $|\Phi_0\rangle$ given by projectors

$$Q_S = |1\rangle\langle 1| + |2\rangle\langle 2| \qquad (14)$$

$$Q_F = |0\rangle\langle 0|, \qquad (15)$$

where the subscripts $S$ and $F$ stand for success and failure, respectively. If one obtains $Q_S$ as a result of measurement, then the state of the system is

$$|\Psi_1^S\rangle = \cos(\theta_1/2)|1\rangle + \sin(\theta_1/2)e^{i\varphi_1}|2\rangle, \qquad (16)$$

which is the original state of the first qubit. The average probability of successful decoding given by the formula

$$q_S = \iint \langle \Phi_0|Q_S|\Phi_0\rangle p(\theta_1,\theta_2|0)d\theta_1 d\theta_2 \qquad (17)$$

is equal to $2/3$. The *a posteriori* probability for preparation $(\theta_1,\theta_2)$ and the results $0$ and $S$ is

$$p(\theta_1,\theta_2|0,S) = \frac{1}{2}\sin^2(\theta_2/2)\sin(\theta_2)\sin(\theta_1). \qquad (18)$$

We see that in the case of successful decoding the a *posteriori* probability is a product of two marginal probabilities $p(\theta_1,\theta_2|0,S) = p(\theta_1|0,S)\,p(\theta_2|0,S)$, where

$$p(\theta_1|0,S) = \int p(\theta_1,\theta_2|0,S)d\theta_2 = \frac{1}{2}\sin(\theta_1) \qquad (19)$$

and

$$p(\theta_2|0,S) = \int p(\theta_1,\theta_2|0,S)d\theta_1 = \sin^2(\theta_2/2)\sin(\theta_2). \tag{20}$$

Moreover, the marginal *a posteriori* probability distribution $p(\theta_1|0,S)$ is equal to the *a priori* probability distribution $p(\theta_1)$. The information on the state of each qubit alone gained in the process of decoding $I_a^D$ can be calculated from the formula

$$I_a^D = -\int p(\theta_a|j)\log_2 p(\theta_a|j)d\theta_a + \int p(\theta_a|j,S)\log_2 p(\theta_a|j,S)d\theta_a. \tag{21}$$

In the case of the decoded qubit (the first qubit in our example) the information gain is $I_1^D = -0.027$ bits and perfectly cancels the information gain $I_1^E$. It means that all the information about the state of the first qubit gained in the process of encoding is erased in the process of decoding. On the other hand, the information gain on the state of the second qubit is $I_2^D = 0.252$ bits. Thus the total information gain on the state of the second qubit is $I_2^E + I_2^D = 0.279$ bits. This is the same amount of information which can be gained in the process of projective measurement given by projectors $|0\rangle\langle 0|$ and $|1\rangle\langle 1|$ performed on the second qubit in the original state $|\Psi_2\rangle$.

It is also worth noting what amount of information we obtain in the case of failed decoding. The calculation similar to the previous one shows that the information gain on the state of each qubit alone is $I_1^F = I_2^F = 0.252$ bits. Thus, the total information on the state of each qubit gained in the processes of encoding and failed decoding is $I_1^E + I_1^F = I_2^E + I_2^F = 0.279$ bits, which is the same amount of information as obtained in the process of projective measurements given by projectors $|0\rangle\langle 0|$ and $|1\rangle\langle 1|$ performed separately on each qubit in its original state.

In conclusion, we have presented a probabilistic scheme which enables to encode the states of two non-entangled qubits in a single qutrit. From this qutrit one can faithfully reconstruct the state of one arbitrarily chosen qubit.

Acknowledgements

We would like to thank the State Committee for Scientific Research for financial support under Grant No. 0 T00A 003 23. One of us (A. G.) would like to thank the European Science Foundation for a Short Scientific Visit under QIT Programme.

*Email address: agie@amu.edu.pl

**Email address: antwoj@amu.edu.pl